\ifdefined\drafting
\documentclass[aps,prl,notitlepage,superscriptaddress,preprint,floatfix,tightenlines]{revtex4-2}
\else
\documentclass[aps,prl,notitlepage,superscriptaddress,reprint]{revtex4-2}

\fi

\ifdefined\drafting
\usepackage[a4paper,left=0.25in,right=1.8in,head=21.0pt]{geometry}
\else

\fi

\ifdefined\showcomments{}
\usepackage[commentmarkup=todo,todonotes={textsize=scriptsize,textwidth=1.6in}]{changes}
\makeatletter \def\@captype{figure} \makeatother 
\else
\usepackage[final]{changes}
\fi

\newcommand{\markhigh}[1]{\bgroup\markoverwith				
	{\textcolor{#1}{\rule[-.5ex]{2pt}{2.5ex}}}\ULon}

\usepackage{scrlayer-scrpage} 

\ifdefined\drafting
\else
\clearscrheadings
\fi

\ifdefined\removefig
\usepackage{comment} 	
\excludecomment{figure}	
	
\else
\fi

\usepackage[utf8]{inputenc} 
\usepackage{siunitx}
\usepackage{graphicx}

\usepackage{amsmath} 
\usepackage{float}

\usepackage{amssymb}
\usepackage{bm} 

\usepackage{hyperref}
\hypersetup{pdfprintscaling=None}

\definechangesauthor[name=Elina Pavlovska, color=violet]{EP}
\definechangesauthor[name=Niels Ubbelohde, color=red]{NU}
\definechangesauthor[name=Martins Kokainis, color=green]{MK}
\definechangesauthor[name=Slava Kashcheyevs, color=blue]{VK}
\definechangesauthor[name=Girts Barinovs, color=brown]{GB}

\newcommand{\beginsupplement}{			
	\setcounter{table}{0}
	\renewcommand{\thetable}{S\arabic{table}}
	\setcounter{figure}{0}
	\renewcommand{\thefigure}{S\arabic{figure}}
	\setcounter{secnumdepth}{2}
}

\newcommand{\taudepx}[1]{[#1]} 

\usepackage{xpatch} 
\makeatletter
\patchcmd{\@ssect@ltx}
{\addcontentsline{toc}{#1}{\protect\numberline{}#8}}
{}
{}
{}
\makeatother

\usepackage{csquotes} 
\usepackage{braket}
\DeclareMathOperator{\sgn}{sgn}

\begin{document}
	\title{Two electrons interacting at a mesoscopic beam splitter}

	\author{Niels Ubbelohde}
	\email[To whom correspondence should be addressed: ]{niels.ubbelohde@ptb.de}
	\affiliation{Physikalisch-Technische Bundesanstalt, 38116 Braunschweig, Germany}
	
	\author{Lars Freise}
	\affiliation{Physikalisch-Technische Bundesanstalt, 38116 Braunschweig, Germany}
	
	\author{Elina Pavlovska}
	\affiliation{Department of Physics,  University of Latvia, 3 Jelgavas street, LV-1004 Riga, Latvia}
	
	\author{Peter G. Silvestrov}
	\affiliation{Institut f\"ur Mathematische Physik, Technische Universit\"at Braunschweig, D-38106 Braunschweig, Germany}
	
	\author{Patrik Recher}
	\affiliation{Institut f\"ur Mathematische Physik, Technische Universit\"at Braunschweig, D-38106 Braunschweig, Germany}
	\affiliation{Laboratory for Emerging Nanometrology Braunschweig, D-38106 Braunschweig, Germany}
	
	\author{Martins Kokainis}
	\affiliation{Department of Physics,  University of Latvia, 3 Jelgavas street, LV-1004 Riga, Latvia}
	\affiliation{Faculty of Computing, University of Latvia, 19 Raina boulevard, LV-1586 Riga, Latvia}

	\author{Girts Barinovs}
	\affiliation{Department of Physics,  University of Latvia, 3 Jelgavas street, LV-1004 Riga, Latvia}
	
	\author{Frank Hohls}
	\affiliation{Physikalisch-Technische Bundesanstalt, 38116 Braunschweig, Germany}
	
	\author{Thomas Weimann}
	\affiliation{Physikalisch-Technische Bundesanstalt, 38116 Braunschweig, Germany}
	
	\author{Klaus Pierz}
	\affiliation{Physikalisch-Technische Bundesanstalt, 38116 Braunschweig, Germany}
	
	\author{Vyacheslavs Kashcheyevs}
	\affiliation{Department of Physics,  University of Latvia, 3 Jelgavas street, LV-1004 Riga, Latvia}

	\begin{abstract}
		The non-linear response of a beam splitter to the coincident arrival of interacting particles enables numerous applications in quantum engineering and metrology yet poses considerable challenge to achieve focused interactions on the individual particle level. 
		Here we probe the coincidence correlations at a mesoscopic constriction between individual ballistic electrons in a system with unscreened Coulomb interactions and introduce concepts to quantify the associated parametric non-linearity. The full counting statistics of joint detection allows us to explore the interaction-mediated energy exchange. 
		We observe an increase from 50\% up to 70\% in coincidence counts between statistically indistinguishable on demand sources, and a correlation signature consistent with independent tomography of the electron emission. Analytical modeling and numerical simulations underpin consistency of the experimental results with Coulomb interactions between two electrons counterpropagating in a dispersive quadratic saddle, and demonstrate interactions sufficiently strong, $U/(\hbar \omega) > 10$, to enable single-shot in-flight detection and quantum logic gates.  
	\end{abstract}

	\maketitle
	

	Coincidence correlations generated by the arrival at separate input ports of a beam splitter have been proven essential for metrology and utilization of single propagating quanta of radiation and matter~\cite{Hong1987, Liu1998, Lopes2015, Toyoda2015, Bocquillon2013}.
	In linear quantum optics these correlations are driven by the  second-order coherence manifested in the Hong-Ou-Mandel (HOM) effect~\cite{Hong1987}, and thereby quantum indistinguishability of the particles can be inferred and certified with extensive  applications in photonic quantum information technologies~\cite{Kok2007, Bouchard2020}. More recently, a fermionic analogue of linear quantum optics has emerged \cite{Bocquillon2012, Bocquillon2013a, Dubois2013, Jullien2014,Baeuerle2018,Takada2019}, leading to quantum tomography of electrical currents \cite{Bisognin2019} via fermionic HOM for well-screened excitations in chiral one-dimensional conductors \cite{Bocquillon2013,Wahl2013,Freulon2015,Marguerite2016,Ferraro2018,Rebora2020}.
	An alternative source of controlled two-particle correlations, pursued for photonic technologies, is direct interaction in the so called quantum non-linear regime~\cite{Chang2014} when the presence of a single quantum sufficiently changes the transparency of the substrate for the other one. While the non-linear response is naturally weak for photons, thus making the engineering of sufficiently strong interactions particularly challenging, the opposite applies to electrically charged fermions, where the natural Coulomb interaction is key to many envisioned applications such as the solid-state flying qubit platform~\cite{Edlbauer2022}.
	Here we present a single-electron quantum optics platform with  on-demand generation and high fidelity detection and demonstrate the measurement of coincidence correlations between individual ballistic electrons at a mesoscopic beam splitter with a picosecond-scale time resolution. We identify generic signatures of the quantum non-linear regime in the first- and second-order correlation signals, supported by quantitative microscopic modelling of the interplay between dispersion of the beam splitter and the Coulomb repulsion. Further we utilize the full counting statistics of joint detection to investigate the interaction-mediated energy exchange between the propagating electrons. Our results demonstrate a mesoscopic beam splitter with Coulomb interactions sufficiently strong for in-flight single-electron detection and potentially a quantum logic gate between time-bin encoded qubits.
	
	\begin{figure*}[htbp]
		\centering
		\includegraphics[width=510pt]{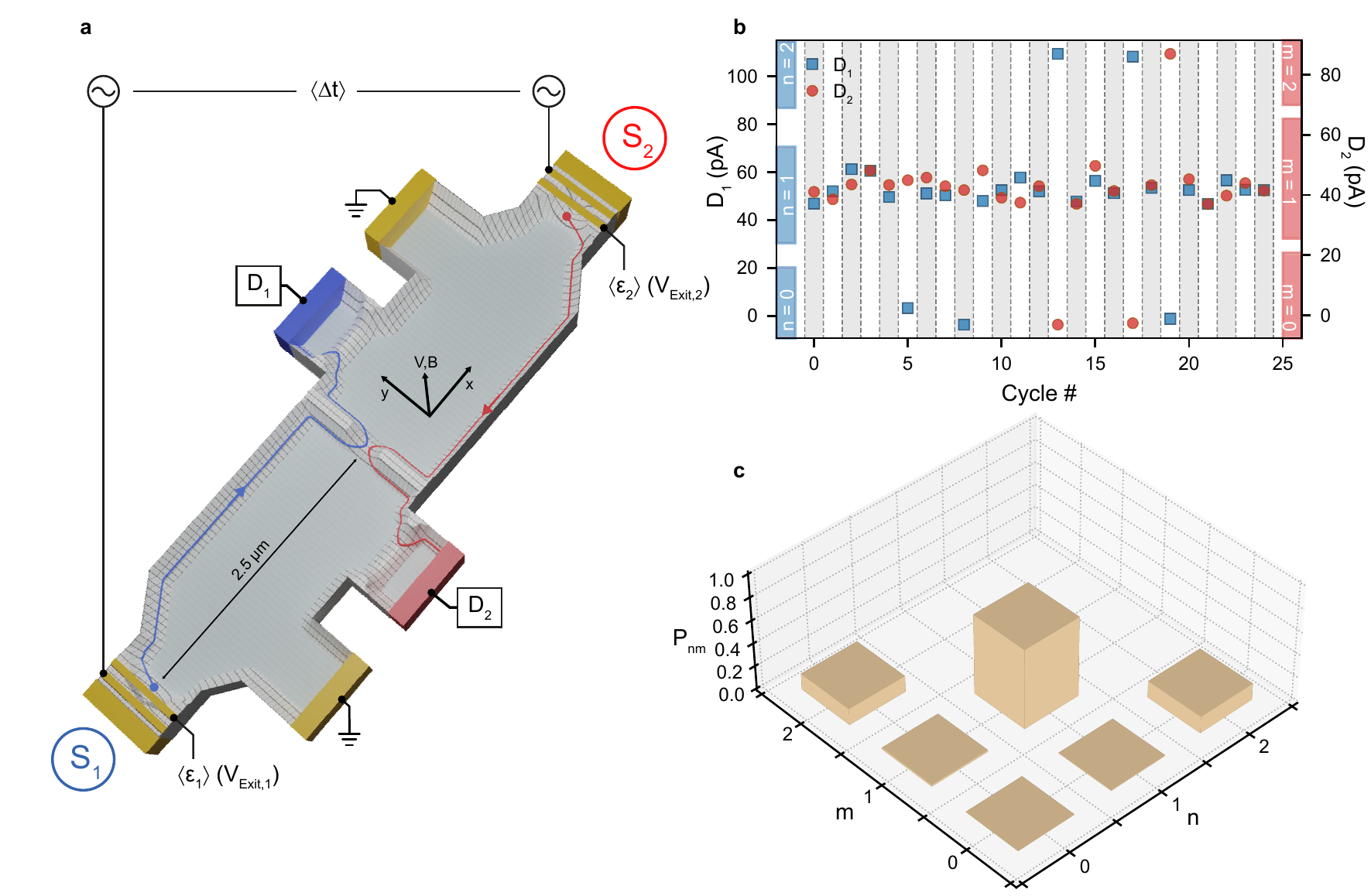}
		\caption{(a) Schematic of the single electron circuit showing the paths (blue, red line) of the two electrons in the transverse confinement potential from the sources (S1, S2) through the beam splitter potential over the entrance barrier towards the single electron detector (D1, D2). Scattered electrons lost to the Fermi sea (blue plane) are removed via ground contacts. (b) Example detector signal for coincident arrival at equal energies. The red/blue bars indicate  \SI{\pm5}{\sigma}-bounds for identification of electron number. (c) The matrix of all counting statistics elements of $P_{nm}$ corresponding to the time trace shown in (b).}
		\label{fig:fig1}
	\end{figure*}
	
	In a HOM-type experiment~\cite{Hong1987}, high time resolution --- not limited by detector bandwidth --- is achieved by precise control of the interarrival time at the beam splitter. In its essential elements the setup is an exact analogue of a two-port scattering experiment and therefore ideally suited to probe interactions utilizing single-particle detectors in this electron quantum optics experiment (for a schematic see Fig.~\ref{fig:fig1}a and Supplementary Fig.~\ref{sec:supp:micrograph}). Two electrons are generated on-demand by tunable barrier non-adiabatic single electron pumps \cite{Kaestner2015}:  a probe electron (2), with fixed average emission energy $\langle \epsilon_2 \rangle$, and a scan electron (1), the energy  of which ($\langle \epsilon_1 \rangle$) is varied. The emission energy and time are well approximated by a correlated \cite{Kashcheyevs2017} bivariate normal distribution $\rho_i(\epsilon_i, t_i)$, parameterized by the widths $\sigma_E$, $\sigma_t$, and the correlation coefficient $r$.  The average emission energy of both electrons is chosen to be some ten meV above the Fermi energy and thereby above the threshold of strong energy relaxation by scattering with the Fermi sea~\cite{Taubert2011,Ota2019,Freise2020} and below the regime of LO-phonon emission~\cite{Fletcher2013, Johnson2018,Emary2019} to ensure low-loss transport. The injected electrons are conducted along the gated mesa edge towards the beam splitter following the chirality imposed by the magnetic field~\cite{Fletcher2013,Ubbelohde2015,Freise2020}.  A notched gate electrode forms a barrier serving as an electronic beam splitter with energy-dependent transmittance $T$, which partitions an incident electron with a  transmission probability $T_i (\langle \epsilon_i \rangle)$. The energy broadening of the transmission threshold is here assumed to be dominated by the energy width $\sigma_{E,i}$ of the energy-time distribution supplied by the source $\rho_i(\epsilon_i,t_i)$, while the barrier is comparatively sharp \cite{Fletcher2019} (see also Methods). The parameters of $\rho_i(\epsilon_i,t_i)$ are inferred for each individual source separately via maximum-likelihood estimation applied to a tomographic protocol utilizing the beam splitter as a tunable energy-filtering barrier \cite{Fletcher2019} (with the respective opposite source switched off).
	The point of half transmission defines the reference levels for the emission energies,  $T_i(0) = 0.5$. The average interarrival time $\langle \Delta t \rangle$ at the beam splitter of the electrons is controlled by changing the emission time $t_i$ of one source with respect to the other, $\langle \Delta t \rangle = \langle t_1 \rangle - \langle t_2 \rangle$. The transmitted and reflected electrons are then trapped via controlled energy relaxation in a detection node and finally read out using a capacitively coupled quantum dot with near unity detection efficiency. A detailed description of this circuit building  block is found in Ref.~\cite{Freise2020}. Repeating the sequence of electron injection, detection, and circuit-level reset builds the counting statistics for the occupancy of each detector node as a result of the partitioning at the beam splitter. A histogram of each detector signal allows to infer a mapping to the corresponding charge transition \cite{Reifert2019, Freise2020}, which then can be applied to the time-trace  to determine the full counting statistics of coincidence correlations $N_{nm}$. The joint detection probability $P_{nm}$ is then estimated as $N_{nm}/\sum_{n,m} N_{nm}$, where $n$ and $m$ are the number of injected electrons reaching detectors $1$ and $2$ respectively. Fig.~\ref{fig:fig1}b and c show an example of measured detector signal time trace and inferred detection probability for coincident arrival ($\langle \epsilon_1 \rangle=\langle \epsilon_2 \rangle=\langle \Delta t \rangle=0$).

	First we focus on lossless partitioning of two colliding electrons, $n=2-m=0,1, \text{ or }  2$, and  $P_{20}+P_{11}+P_{02}=1$. We characterize coincidence correlations by  the first and the second order correlation signals,  $s^{(1)}$ and $s^{(2)}$, defined as deviations from the uncorrelated baseline of the intensity difference, $ (\langle n \rangle -\langle m \rangle)/2= P_{20}-P_{02}$ and the interbeam intensity correlation  $\langle n \, m\rangle=P_{11}$, respectively.
	The uncorrelated baseline expectation for partitioning probabilities at sufficiently large arrival time mismatch is set by addition of statistically independent transmission-or-reflection events, $P_{11}=T_1 \ T_2 + (1-T_1)(1-T_2)$ and $P_{20}-P_{02}=T_1-T_2$. 
	As the probe electron is always kept at  $\langle \epsilon_2\rangle\!=\!0$
	regardless of the scan electron tuning,
	the correlation signals for our special case $T_2=0.5$ are
	\begin{align} 
	s^{(1)}  & = P_{20}-P_{02} -T_1 +T_2 \, , \label{DeltaI} \\
	s^{(2)}  & = P_{11}-0.5 \, . \label{DeltaC} 
	\end{align}
	
	In linear quantum optics, first order quantities are unaffected by coincident arrival  ($s^{(1)}=0$ for any $T_{1,2}$), but
	quantum mechanical interference of two-particle amplitudes modifies the antibunching probability by the overlap of wave-functions of  the transmitted ($\ket{\psi_T}$) and the reflected ($\ket{\psi_R}$)  particles, $P_{11} \to P_{11}\pm 2 \, |\langle \psi_R \vert \psi_T \rangle |^2$ with $-$ ($+$) sign corresponding to boson (fermion) exchange statistics (the HOM effect).
	However, for the single-electron sources in this experiment the estimated upper bound on the visibility of this HOM effect in $s^{(2)}$, $\mathrm{Tr}\left(\hat{\rho}_1 \hat{\rho}_2 \right) \simeq 0.07$, is small.  Conversely, the essence of non-linear correlations can be understood qualitatively as mutual gating, if the presence of one electron at the beam splitter changes the transmission probability for the other. Yet without a microscopic model, interpretation of such conditional probabilities may be problematic, since partitioning events are not statistically independent in general, and a coarse mean-field approximation would average out possible dependence of the gating strengths on the relative arrival time.
	
	\begin{figure*}[htbp]
		\centering
		\includegraphics[width=510pt]{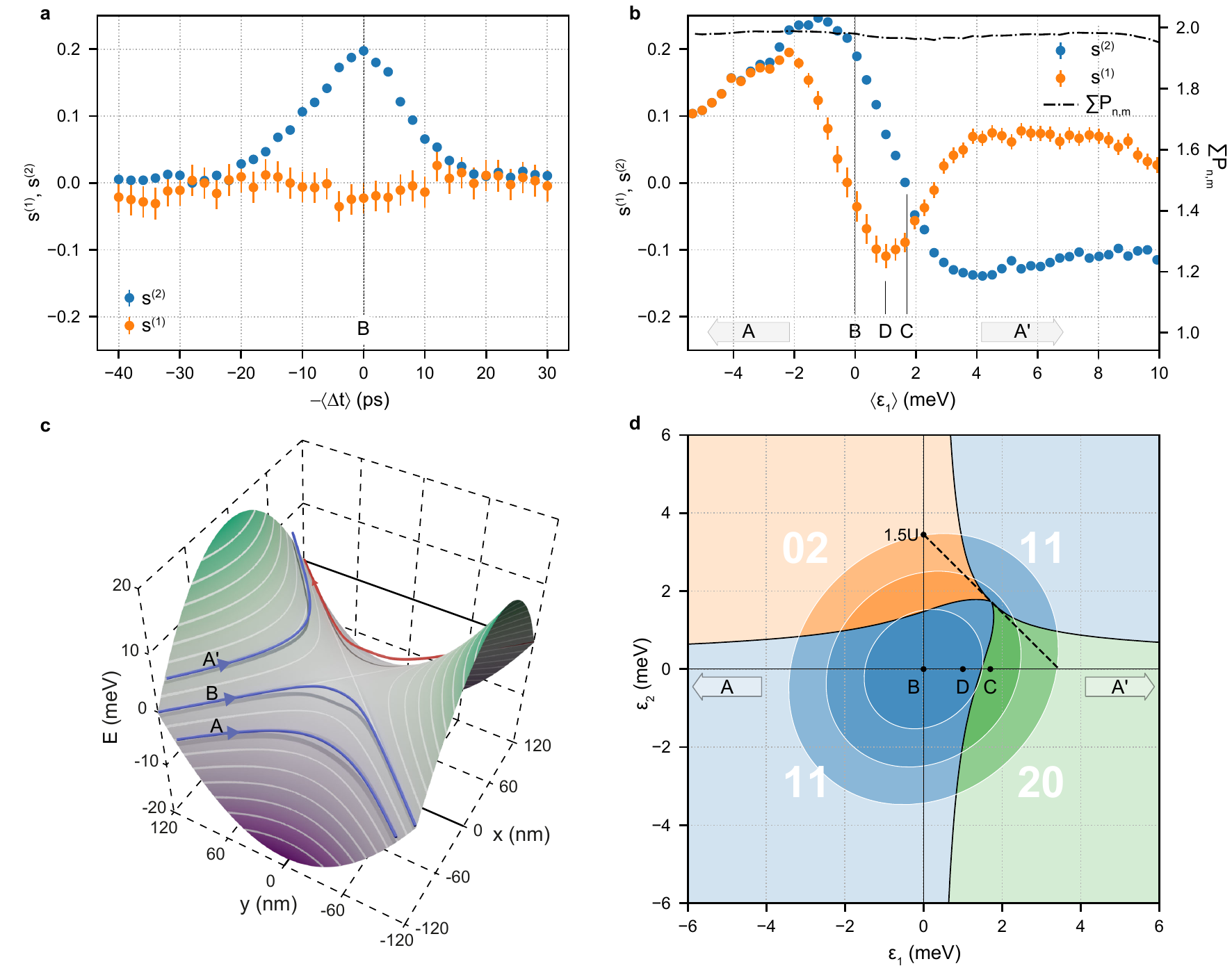}
		\caption{First and second order correlation signals (a) for $\langle \epsilon_1 \rangle = \langle \epsilon_2 \rangle = 0$  and (b) for $\langle \Delta t \rangle = 0, \langle \epsilon_2 \rangle = 0$ together with the average number of electrons jointly detected. (c) The colored lines show individual electron trajectories in the horizontal plane of two spatial dimensions with the height indicating the potential experienced by the scan (blue) and the probe (red) electrons for three example initial conditions, corresponding to points A, B and A' in (b). The surface and the associated level lines show the external guiding potential with the dark gray shadows highlighting the external potential contribution to the total potential guiding the motion of the interacting electrons. (d) Schematic  showing the transmission thresholds, $\epsilon_1^{\ast}(\epsilon_2)$ and  $\epsilon_2^{\ast}(\epsilon_1)$,  and the resulting partitioning (colored domains) for  $\Delta t = 0$. The ellipses mark the joint probability distribution of $\epsilon_1$ and $\epsilon_2$ for pairs emitted at $\Delta t = 0$ in case B ($\langle \epsilon_i\rangle=0$);  contour levels are chosen at intervals equivalent to the coverage factor of a normal distribution.
			\label{fig:fig2}}
	\end{figure*}
	
	The clear definition of mechanism-independent signals allows us now to probe the time-width of coincidence correlations for conditions, where the scan source is tuned most similar to the probe source and the main difference is set by a non-zero interarrival time $\langle \Delta t \rangle $.  In Fig.~2a, $s^{(2)}$ shows a clear correlation signature with a peak amplitude of 0.2 and the second moment width of $\sigma^{(B)}_t=\SI{20}{ps}$, while $s^{(1)}$ does not a show a discernible deviation from zero within the measurement uncertainty. This result is seemingly consistent with second-order coherence between sources of indistinguishable particles but on its own does not rule out interaction-driven correlations. Indeed, as defined by Eqs.~\eqref{DeltaI}-\eqref{DeltaC},  $s^{(2)}$ is even and $s^{(1)}$ is odd  w.r.t.\  the exchange of the two electrons. If the two  sources (which are independent by design) generate  indistinguishable energy-time distributions, then $\langle \Delta t \rangle \to -\langle \Delta t \rangle$ is equivalent to   $1 \leftrightarrow 2$, and $s^{(1)}$ must vanish at $\langle \Delta t \rangle=0$. Whether $s^{(1)} \not =0$ is resolved as function of $\langle \Delta t \rangle $ is determined by the competition between the symmetry-breaking effect of arrival time difference and the reduction of the interaction strength at large $ | \langle \Delta t \rangle |$. The experimental results for the $T_1=T_2=0.5$ setting show  that electrons are insufficiently distinguished in their transmission properties by the average difference in emission time alone.
	
	To reveal the unambiguous signature of interaction-induced correlations, we break the symmetry by tuning the average emission energy $\langle \epsilon_1 \rangle $ of the scan electron while keeping the probe electron at half transmission. In Fig.~2b, also the first-order  signal now clearly presents a correlation signature, $s^{(1)} \not = 0$ as a function of $\langle \epsilon_1 \rangle$. In the following, specific key features of both signals, labelled A--D in Fig.~\ref{fig:fig2}b, will be discussed to build a generic qualitative picture of the correlation-generating mechanism.
	
	In regions A and A$'$ the energy $\langle \epsilon_1 \rangle$ is so far detuned, that the scan electron is always reflected (A: $P_{20}=0$) or always transmitted (A$'$: $P_{02}=0$), and as a consequence, $s^{(2)}=+ s^{(1)}$  or $-s^{(1)}$, respectively. Equation \eqref{DeltaI} with $T_1=0$ or $1$ suggests interpreting $\mp s^{(1)}=T_2^{\ast}-T_2<0$ as a non-linear modification of the probe electron transmission function, quantifiable by the shift of the corresponding half-transmission threshold from $0$ to $\epsilon_2^{\ast}>0$ where  $T_2^{\ast}=T_2(\epsilon_2^{\ast})$.
	This simple heuristic can be generalized to a quantitative statistical model of two-electron collision, if the randomness in partitioning comes predominantly from the spread of the incoming energy and arrival time distributions $\rho_i(\epsilon_i, t_i)$.  We introduce two functions,
	$\epsilon_2^{\ast}(\epsilon_1, \Delta t)$ and 
	$\epsilon_1^{\ast}(\epsilon_2, \Delta t)$, which, plotted jointly in the $( \epsilon_1, \epsilon_2)$ plane for a fixed value of $\Delta t$, define four  domains corresponding to definite scattering outcomes: transmission ($\epsilon_i> \epsilon_i^{\ast}$) or reflection ($\epsilon_i< \epsilon_i^{\ast}$) of the electron incoming from the source $i$ with a well-defined  energy $\epsilon_i$ and interarrival time $\Delta t$. A schematic of such four-fold partitioning of the initial conditions is shown in Figure~\ref{fig:fig2}d for the time slice $\Delta t=0$. Subsequently, the counting statistics is obtained by integrating over the corresponding domains,
	\begin{align} \label{eq:genstatistics}
	(P_{20}, P_{02})=\int \prod_{i=1,2} d \epsilon_i \, d t_i \, \Theta[\pm \epsilon_i\mp \epsilon_{i}^{\ast}(\epsilon^{}_{\bar{\imath}}, t_1-t_2)] \, \rho_i(\epsilon_i ,t_i) \, ,
	\end{align}
	where the upper sign is for $P_{20}$,
	$\bar{\imath}=3-i$ is 
	the electron index opposite to $i$. Here Heaviside step function $\Theta$ approximates a sharp transmission function of negligible quantum width in energy. 
	
	For an inversion-symmetric beam splitter and simultaneous arrival ($\Delta t =0$), the shifted thresholds have to be symmetric with respect to exchange of the sources, $1 \leftrightarrow 2$,  hence both can be described by one single-argument function, $\epsilon_i^{\ast}(\epsilon, 0)=\epsilon^{\ast}(\epsilon)$. The symmetric crossing point $\epsilon^{\ast}(3U/4)=3U/4$ defines a characteristic energy $U$ of mutual gating.
	Having introduced the generic  transmission threshold shifts as quantifiers of non-linearity in 
	Figure~\ref{fig:fig2}d and Eq.~\eqref{eq:genstatistics}, we can explain the remaining marked features of the experimental signals in  Figure~\ref{fig:fig2}b. 
	At B, the joint distribution symmetrically contributes to $P_{20}$ and $P_{02}$, and the difference in $s^{(1)}$ will therefore vanish, whereas at C, $\epsilon_1$ is sufficiently large compared to $\epsilon^{\ast}$  for only \SI{50}{\%} of the incident electron pairs to be jointly deflected, indicated by the zero crossing of $s^{(2)}$. When the joint distribution mostly probes the difference between the zero shifted and shifted threshold at positive energies, $0< \langle \epsilon_1 \rangle <\epsilon_1^{\ast}(0)$ (the area between $\epsilon_2=\epsilon^{\ast}_2$ and $\epsilon_2=0$ in Fig. 2d), $s^{(1)}$ will assume its most negative value (D). The mutual gating captured by the shifted-threshold function $\epsilon^{\ast}(\epsilon)$ thus offers qualitatively a complete picture of both correlation signals at $\langle \Delta t\rangle=0$, but does not quantify the scaling of the thresholds with $\Delta t$, nor, without binding $\epsilon^\ast$ to a specific model, predicts the dependence on $\sigma_E, \sigma_t$ of the sources.
	
	\begin{figure*}[htbp]
	\centering
	\includegraphics[width=510pt]{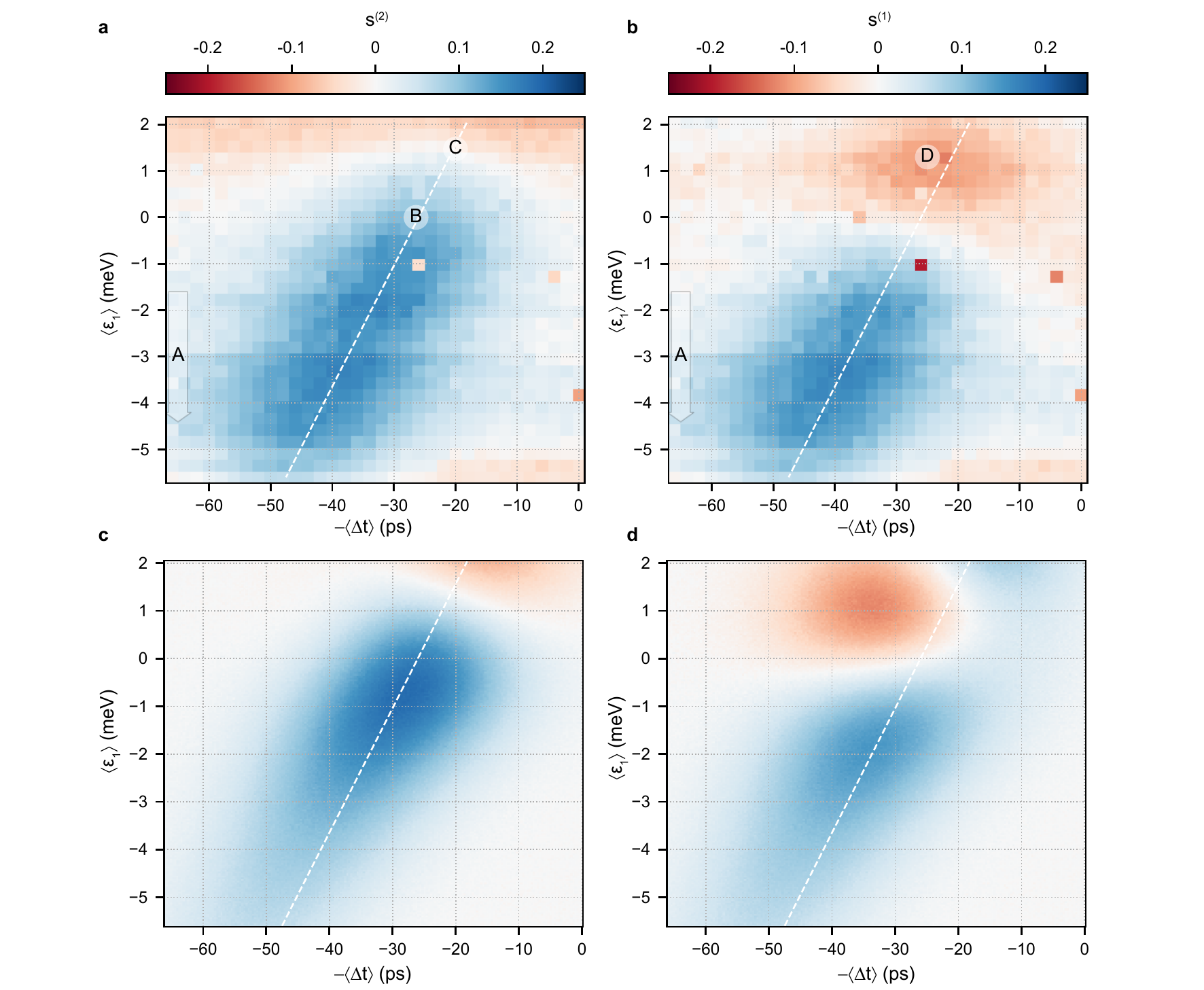}
	\caption{(a),(b) Measured  and (c),(d) simulated  correlation signals $s^{(2)}$, $s^{(1)}$. For $s^{(1)}$, $T_1(\langle \epsilon_1 \rangle) - T_2(\langle \epsilon_2 \rangle)$ is estimated at large $|\langle\Delta t\rangle|$ and subtracted from $P_{20} - P_{02}$. Simulation parameters: $U=\SI{2.3}{meV}$, $\omega=\SI{250}{GHz}$, $d_\text{cutoff} = \SI{75}{nm}$ and $\omega_x/\omega_y = 1$. The white, dashed, straight line indicates the expected energy-time correlation of the source emission, offset to coincide with the maximal joint detection probability $P_{11}$ at $\langle \epsilon_1 \rangle=0$, which also defines $\Delta t = 0$ for the simulation. The labels identify key features discussed in the main text.}
	\label{fig:fig3}
\end{figure*}	
	
	To compute the threshold functions and interpret the decay of the correlation signals, we introduce a microscopic model. The separation of scales between the fast cyclotron motion localized into the lowest Landau state and $\bm{E} \times \bm{B}$ drift \cite{Kataoka2016} of the corresponding guiding centres along smoothly varying equipotential lines~\cite{Ryu2016} allows application of the classical equations of motion for two interacting electrons~\cite{Pavlovska2022}, once the beam splitter $V(x_i,y_i)$ and the interaction $V_{\text{int}}(r)$ potentials are specified. A versatile analytical model is enabled~\cite{Pavlovska2022} by approximating the beam splitter potential with a quadratic saddle~\cite{Fertig1987, Buttiker1990, Pavlovska2022, Ryu2022}, as defined in Methods and summarized below. First, the symmetric threshold function $\epsilon^{\ast}(\epsilon)$ is implicitly expressed via the interaction-induced change of the relative coordinate traversal time $\Delta \tau(E)$ 
	\begin{equation}
	\left(\epsilon^{\ast}/\epsilon\right)^{\pm 1} = 
	\tanh [ \omega \Delta \tau( \epsilon^{\ast}+\epsilon)/2] \, , \label{eq:threshold}
	\end{equation}
	where the `$+$'  sign is for  $\epsilon^{\ast} < |\epsilon|$. Equation \eqref{eq:threshold} reflects the competition \cite{Ryu2022,Pavlovska2022} of the energy-dependent interaction effect encoded in $\Delta \tau(E)$ and the linear dispersion of the beam splitter characterized by $\omega=m \omega_x \, \omega_y/(e B)$. 
	Second, the $\Delta t $ dependence is obtained from Eq.~\eqref{eq:threshold} by scaling relations~\cite{Pavlovska2022}  $\epsilon_{i}^{\ast} (\epsilon_{\bar{\imath}}^{}, \Delta t)= 2 \epsilon^{\ast}([1+e^{\pm \omega \, \Delta t}] \epsilon_{\bar{\imath}} /2) / (1+e^{\mp \omega \, \Delta t})  $ with the upper sign for $i=1$.  These relations obey the symmetry-dictated invariance under $\Delta t \to -\Delta t$ and $1\leftrightarrow 2$.  In particular, the  crossing point of the thresholds stays on a straight line (dashed line in Fig.~\ref{fig:fig2}d), $\epsilon_{1c}^{\ast}+\epsilon_{2c}^{\ast}= 1.5 U$  with late-to-arrive electron experiencing a stronger gating effect, $\epsilon_{1c}^{\ast}>\epsilon_{2c}^{\ast}$  for $\Delta t>0$.

	To characterize the extent of the correlation signatures in relative energy and time, in Fig.~\ref{fig:fig3} both $\langle \epsilon_1 \rangle$ and $\langle \Delta t \rangle$ are scanned for a modified operating point of the sources with reduced driving amplitude. In addition to reproducing the key features identified and labelled in Fig.~\ref{fig:fig2}, these results illustrate how the correlation signatures both in $s^{(1)}$ and $s^{(2)}$ persist over a wide range of energies beyond the energy width of the source distributions, covering both zero and full transmission of the scan electron.  The width of the correlation signal  as function of $\Delta t$ remains fairly constant; in region A at $\langle \epsilon_1 \rangle=-\SI{2.8}{meV}$ we measure $\sigma^{(A)}_{t} =\SI{14}{ps}$. Subtracting the broadening due to source distributions and using the perturbative limit of Eq.~\eqref{eq:threshold} yields an estimate of $\omega=\SI{270}{GHz}$ (see Methods).
	Taking the maximal joint detection probability $P_{11}=s^{(2)}+0.5$ to be occurring close to coincident arrival, the apparent shift of this point with $\langle \epsilon_1 \rangle$ should follow the energy-time correlation $r \sigma_t/\sigma_E$ of the source, since for tunable barrier pumps the emission energy is linked to the emission time via the drive signal modulating the dynamic quantum dot (the correlation coefficient $r$ is a fingerprint of the same mechanism \cite{Kashcheyevs2017}). The predicted time shift (white dashed line) agrees well with observed correlation signal.

	In Fig.~\ref{fig:fig3}c and d, the experimental results are compared to a simulation using a Coulomb potential with adjustable parameters $\omega$,  $\omega_x/\omega_y$ and $U$ (which leaves the effective charge as a free parameter). To account  for both the limitations of the quadratic  saddle approximation and screening, we neglect correlations beyond a  distance $d_{\text{cutoff}}$  between the electrons, see Methods. Magnitude and extent of both positive and negative correlations are well reproduced for both first and second order signals, including the positive excess energy required to overcome gating by the second electron (C in $s^{(2)}$), as well the large area of the interaction-driven correlation signal in region A.
	Quantitatively, the estimated parameters are in reasonable agreement with unscreened Coulomb interactions  ($U=\SI{2.3}{meV}$ versus $U=\SI{2.6}{meV}$, see Supplementary Note~\ref{sec:supp:powerlawpotential}) near the centre of the beam splitter.  The large ratio $U/(\hbar \omega) =14$ justifies the semiclassical approach to interactions \cite{Pavlovska2022} and confirms that the experiment is in the interaction-dominated regime \cite{Bellentani2019}.
	The simulations also explain the unresolved sign of the interarrival time for statistically equal energy collisions, $s^{(1)}(\langle \Delta t \rangle)\approx 0$ for $\langle \epsilon_i \rangle=0$ in Figures \ref{fig:fig1}a, \ref{fig:fig3}b and \ref{fig:fig3}d, as the dispersive delay $\sim \omega^{-1} \ln (d_{\text{cutoff}}/d_{\text{min}})$, cut off by the finite size of the beam splitter, limits the dynamical range required to cover the full  emission time distribution. The discrepancies in the comparison with the experimental data, such as the excessive shift of $s^{(1)}$ minimum (point D) with respect to the inferred coincidence line and the faster decay of the $s^{(1)} \approx s^{(2)}$ signal for $\langle \epsilon_1 \rangle \lesssim-\SI{3}{meV}$, 
	are likely to stem from microscopic details beyond the idealizations of the model.
	
	\begin{figure*}[htbp]
		\centering
		\includegraphics[width=510pt]{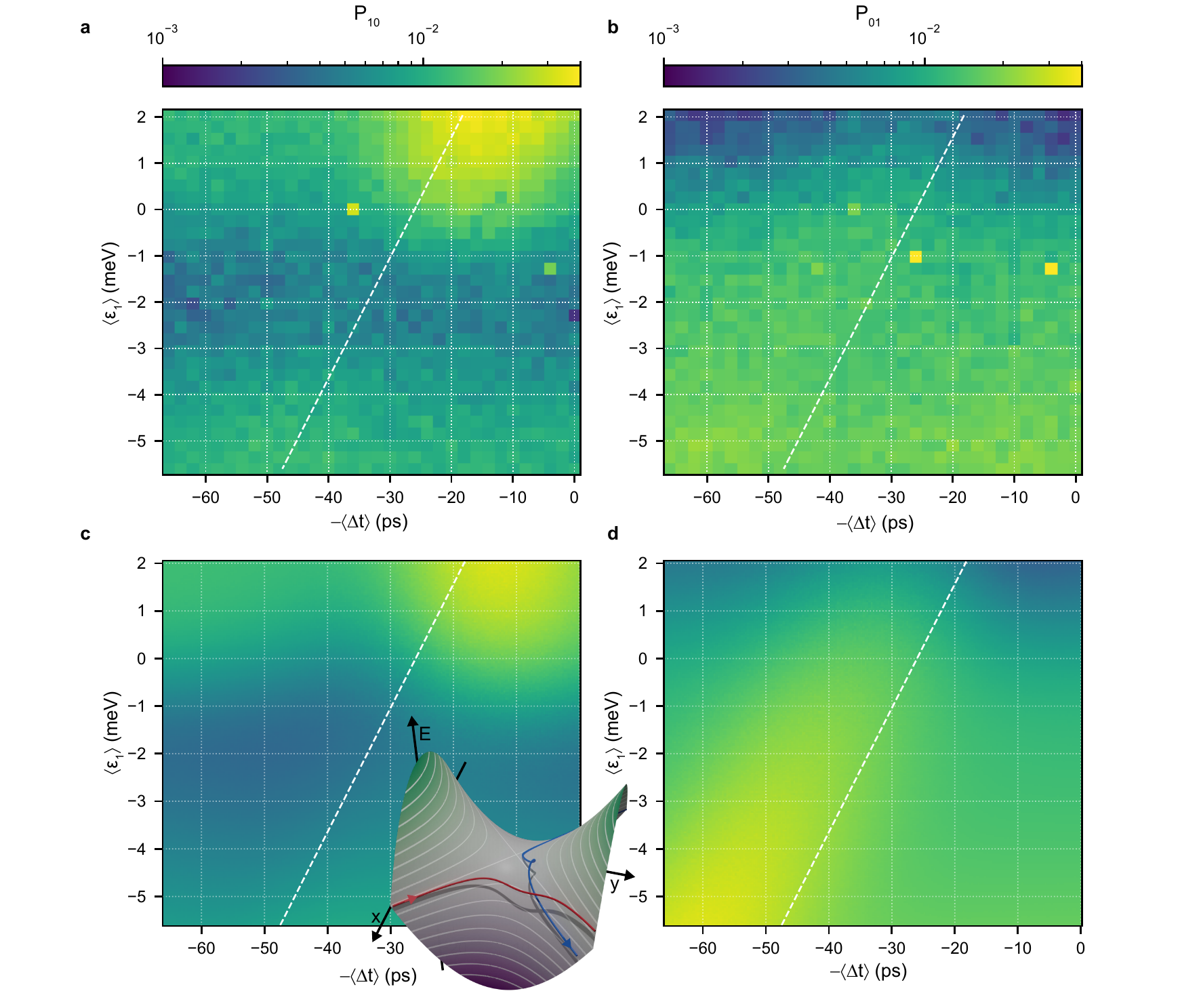}
		\caption{Detection probabilities, (a) $P_{10}$ and (b) $P_{01}$, corresponding to the acquisition cycles, where one electron was lost before reaching the detector. (c), (d) Simulation of $P_{10}$ and $P_{01}$ taking into account the elastic energy exchange with parameters as in Fig.~\ref{fig:fig3} and an energy-dependent loss rate during propagation after the collision, as detailed in Supplementary Note \ref{sup:energy_exchange}. The white dashed line again indicates estimated coincident arrival. The inset shows example trajectories with $\epsilon_1 = \SI{0.1}{meV}$ and $\Delta t = \SI{17}{ps}$ leading to large energy exchange using the same style as in Fig.~\ref{fig:fig2}c.}
		\label{fig:fig4}
	\end{figure*}
	
	In addition to coincidence correlations, access to the full counting statistics also allows to analyze cases where less than two electrons reach the detectors (see Fig.~\ref{fig:fig1}c).  For the injected electrons, any energy loss will drastically increase the rate of scattering with electrons from the Fermi sea during propagation \cite{Ota2019, Freise2020} and thus make it unlikely for the electron to pass over the barrier at the entrance towards the detection node (see also Fig.~\ref{fig:fig1}a). Rare cases where one electron is lost are counted in $P_{10}$ and $P_{01}$ in Fig.~\ref{fig:fig4}. Both signals show a $\langle \Delta t \rangle$-independent background, which reflects the energy dependence of loss rate and transmission probability, as it is more likely  to find at least one electron at detector 1 (enhanced $P_{10}$) instead of detector 2 (reduced $P_{01}$) for positive $\langle \epsilon_1 \rangle$. The most striking feature in $P_{10}$ is the appearance of enhanced losses near the coincidence line, which cannot be explained by the concept of shifted transmission thresholds alone, yet will be confirmed by the microscopic model.  The analytic computation of the energy exchange between the two electrons (see Supplementary Note \ref{sup:energy_exchange}), which together with an approximation of the energy-dependent loss rate after the collision yields a prediction for $P_{10}$ and $P_{01}$ shown in Fig.~\ref{fig:fig4}c and d using the simulation parameter set from Fig.~\ref{fig:fig3}. Based on the good agreement, we identify the enhanced signal in $P_{10}$ as explicit evidence for interaction-mediated energy exchange at the beam splitter. In the saddle approximation, the energy exchange is maximized, when $\langle \epsilon_1 \rangle$ and $\langle \Delta t \rangle$ are just large enough to allow electron 1 to be transmitted and then trail behind electron 2 (see inset in Fig.~\ref{fig:fig4}c). A complimentary signal of energy exchange in $P_{01}$ is more difficult to discern above the noise. Due to the large $\langle \epsilon_1 \rangle$ and $\langle \Delta t \rangle$ offsets in the latter case, interactions in the experiment are likely attenuated by screening.
	An additional fingerprint of energy exchange at the beam splitter can be observed in $P_{00}$ (see Supplementary Fig. \ref{sec:supp:additionaldata}), where a small but clearly discernible signal traces the coincidence line. Loss of both electrons might be induced by sufficiently large  energy  transfer in the elastic collision such that subsequent LO phonon emission  becomes relevant, yet other, unexplored, inelastic loss mechanisms may also be present.
	
	The results above demonstrate how precise coincidence counting measurements of correlation signatures can be used to explore interactions between individual electrons colliding at a beam splitter and to quantify the non-linear response  by the shift of the transmission threshold --- a measure of parametric non-linearity at a single-quantum level.  The capability for the energy spectroscopy of the scattering outcomes promotes further experimental  studies. While in linear optics the HOM effect enables metrology and tomography by being time-sensitive  to the overlap of two impinging wave packets, in the non-linear regime $U/\hbar \gg \omega  $, interactions at the  beam splitter provide a classical counterpart in measuring the statistical indistinguishability between independent sources with $\omega \gg \sigma_t^{-1}$. Here the dispersion of the beam splitter $\omega$ emerges as a crucial scale for metrology: large $\omega$ boosts the sensitivity to interarrival times, while small $\hbar \omega$ is essential for the selectivity of energy-time filters \cite{Locane2019,Fletcher2019}. The width of the emission-time distribution  measured here exhibits consistency between the two tomography techniques.
	
	Furthermore, the  demonstrated interaction strength could be exploited in quantum information technology as follows. An electron travelling ballistically  in a superposition of two short wave-packets spaced at distances of multiple $\omega^{-1}$ forms a time-bin encoded qubit which can experience a logic-state dependent phase rotation by $\Delta \varphi$ from the other electron counterpropagating on the opposite side of the beam splitter. To estimate feasibility of quantum logical operations, $\Delta \varphi  \sim \pi$, consider coincident arrival with $\epsilon_1 =\epsilon_2=\epsilon/2<0$, 
	such that both electrons are reflected. The phase shift accumulated along the classical trajectory 
	in the perturbative limit of Eq.~\eqref{eq:threshold} evaluates exactly to
	\begin{align}
	\Delta \varphi = \hbar^{-1} \int V_{\text{int}}[ r(t)]  dt =  \frac{\epsilon^{\ast}(\epsilon)}{ \hbar \omega} \label{eq:figureotmerit}
	\end{align}
	without violating the Heisenberg uncertainty for the wave-packet time-energy localization if $U \gg \hbar \omega$  (see Supplementary Note \ref{sup:interaction_strength}). 
	$\Delta \varphi$ of order one is possible as
	$\hbar \omega \sim \epsilon^{\ast}(\epsilon) \ll U$ is the classical perturbative limit condition.
	Using the parameter range from Fig.~\ref{fig:fig3} where the analytic model has been validated with $\epsilon^{\ast}(\SI{-2.8}{meV})=\SI{0.88}{meV}$, we estimate $\Delta \varphi=1.7 \pi$, supporting the potential for quantum applications.
	Equation \eqref{eq:figureotmerit} is also an important figure of merit for sensing, e.g. single-shot readout of the flying qubit described above: shift of the threshold of a probe electron by $ \epsilon^{\ast}$ can be efficiently detected by scattering if it exceeds the quantum width $\sim \hbar \omega$  of the transmission function. The above estimates show that efficient exploitation of interactions requires strong non-linearity,  $U/(\hbar \omega) >1$, as $\epsilon^{\ast}(\epsilon)$ is at most $\sim U$.


	\section*{Methods}
	\subsection*{Experimental realization}
	
	The circuit has been realized  \cite{Gerster2018} using a GaAs/AlGaAs heterostructure with \SI{97}{\nano\meter} nominal 2DEG depth, charge carrier density $\SI{1.9e11}{\per\square\centi\meter}$, mobility $\SI{1.15e6}{\square\centi\meter\per\volt\per\second}$, and quantum life time $\SI{2.9}{\pico\second}$. The circuit elements are formed by an shallow-etched channel and Cr/Au gate electrodes. Measurements were performed in a dilution refrigerator with base temperature below \SI{50}{\milli\kelvin} and at $B=\SI{10}{\tesla}$ perpendicular magnetic field  (filling factor $\nu\approx 1$). Electrons are generated on-demand by driving the single electron pumps with a single period of a \SI{300}{MHz} sine signal. The interarrival time is tuned by adjusting the channel skew between the signal sources. A depletion gate \cite{Kataoka2016,Johnson2018} over the mesa edge of the wave guide is set to \SI{-200}{mV} resulting in a transverse confinement energy $\hbar\omega_y$ at the mesa edgee of \SI{2.7}{meV}, estimated from the energy dependence of LO phonon emission \cite{Emary2019}. For the electron propagating along the path towards the beam splitter, the dispersion in the transverse confinement potential adds an additional contribution to the energy-time correlation of the source shown and discussed in Fig.~\ref{fig:fig3} and Fig.~\ref{fig:fig4}, but which, with a coarse estimate of $\sim \SI{4}{ps}$, can be considered negligible relative to the source contribution. The dispersive beam splitter potential induced by a notched gate finger (see also Supplementary Fig.~\ref{sec:supp:micrograph}) can be characterized in time-of-flight measurements \cite{Kataoka2016} yielding an order of magnitude estimate of $\omega=\SI{300}{GHz}$ \cite{Pavlovska2022} for comparable gate geometries. The parameter estimates for two source distributions  
	$\rho(\epsilon, t) = \rho(\sigma_E \tilde{\epsilon}, \sigma_t\tilde{t}) = $
	$1/\left(2 \pi \sqrt{1-r^2}\right) \exp \left[ -\left(\tilde{\epsilon}^2 - 2 r \tilde{\epsilon} \tilde{t} + \tilde{t}^2\right)/\left(2 (1-r^2)\right)   \right]$
	are  $\{\sigma_E,\sigma_t,r\}$ = $\{\SI{1.08}{meV}, \SI{8.3}{ps}, 0.50\}$ for the source electron and $\{\SI{1.07}{meV}, \SI{6.9}{ps}, 0.52\}$ for the probe electron. Here $\epsilon$ and $t$ are statistical deviations from the central averages, $\langle \epsilon_i \rangle$ and  $\langle t_i \rangle$.
	

	\subsection*{Analytical  model}
	For analytical calculations we use quadratic saddle approximation \cite{Fertig1987} of the single-particle beam splitter potential $V(x_i,y_i)=m(\omega_y^2 y_i^2-\omega_x^2 x_i^2)/2$, where $m$ is the effective mass. Recent work \cite{Pavlovska2022} has shown that in this case, besides the total energy $\epsilon_1+\epsilon_2$, an additional quantity  $\mathcal{H} =\epsilon_1'+\epsilon_2'$ with $\epsilon_i'=(1+e^{\mp \omega \Delta t})\epsilon_i/2$ (upper sign for $i=1$) is conserved in collisions,  for any interaction potential $V_{\text{int}}$ decaying sufficiently fast within the range of validity of the quadratic saddle approximation. This extra conservation law corresponds to equipotential drift of the relative coordinates $x=x_2-x_1$, $y=y_2-y_1$ generated by the Hamiltonian flow, $\mathcal{H}(x,y) = V(x,y)/2 + V_{\text{int}} \left(\sqrt{x^2+y^2}\right)$ such that $p = e B y$ is the canonical momentum, conjugate to $x$~\cite{Pavlovska2022}.  The effects of interactions relevant for the computation of shifted thresholds and the energy exchange are conveniently encoded in terms of  $\Delta \tau(E) = \tau(E) - \tau_0(E)$
	where  $\tau= \int |dx/(\partial \mathcal{H}/\partial p)| $ is total traversal time for the relative coordinate, computed along the trajectory $\mathcal{H}(x,y)=E$ implicitly defined by the conservation law, and $\tau_0$ denotes $\tau$ in the absence of interactions ($V_{\text{int}}=0$ identically). 
	$\Delta \tau(E)$ diverges at  $E=3 U/2$ which is the minimal $E$ required for the two electrons to pass each other (i.e., reach $x=0$), thus setting the scale for the magnitude of the mutual gating effect. 
	For the unscreened Coulomb potential, $V_{\text{int}}(r)=U \, d_0 /r$, we derive a closed form expression for $\Delta \tau(E)$ which is then used to determine the thresholds via Eq.~\eqref{eq:threshold} and compute the final energies after the collision. 
	In simulations, the finite size of the beam splitter region is accounted for by setting  $\Delta \tau(E)=0$ for $|E|> U_c^{\pm}$ where the  thresholds $U^{\pm}_c$ are chosen by fixing an interaction cut-off distance 
	$d_{\text{cutoff}}$. Derivation details are given in the Supplementary Notes \ref{sec:app:DeltaTau},   \ref{supp:thresholdtau}, \ref{sec:supp:cutoffimplem}, and
	\ref{sup:interaction_strength}.

	\subsection*{Perturbative limit for the non-linear shifts and \texorpdfstring{$\omega$}{omega} estimation\label{sup:perturbative}}
	Under the condition  $\epsilon^{\ast} \ll |\epsilon| $ Eq.~\eqref{eq:threshold} simplifies to  $\epsilon^{\ast} = \epsilon \, \omega \, \Delta \tau(\epsilon)/2 = c_{\pm}  V_{\text{int}}(d_{\text{min}})$ where $d_{\text{min}}(\epsilon)$ is the closest approach distance between particles and $c_{\pm}$ is a constant of order $1$ which depends on the exact shape of the potentials.  We derive the exact prefactors for the case of $V_{\text{int}}(r) \propto r^{-\alpha}$ to prove Eq.~\eqref{eq:figureotmerit} and relate the tail of the interaction potential to that of the shifted threshold, $\epsilon^{\ast}(\epsilon) \propto  (U/|\epsilon|)^{\alpha/2} U$, see Supplementary Notes~\ref{sec:supp:powerlawpotential} and \ref{sup:interaction_strength}. The latter relation captures the contribution of the increasing distance $d_{\text{min}}$ ($\propto | \epsilon|^{-1/2}$ in the quadratic saddle) and the decay of the interaction potential.
	Further applying the scaling relations, we get
	$\epsilon_2^{\ast} \propto (1+e^{- \omega \, \Delta t})^{-\alpha/2} (1+e^{+\omega \, \Delta t})^{-1}$ for constant $|\epsilon_1| \gg   \epsilon^{\ast}_2$.
	The second moment of  the latter $\Delta t$-dependence is $(\sigma_t^{\text{min}})^2= [\pi^2/6+ \Psi'(\alpha/2)] \, \omega^{-2}$ where $\Psi$ is the digamma function. 
	This gives $\omega = (2/3)^{1/2} \pi /\sigma_t^{\text{min}}$ for the Coulomb ($\alpha=1$) potential.
	
	
	This is used to assess sensitivity of the probe electron for small gating effect, $\epsilon^{\ast}_2 \lesssim \sigma_{E2}$, and large energy detuning of the scan electron, $|\langle \epsilon_1 \rangle | \ll \sigma_{E1}$, as $s^{(1)}=s^{(2)}\propto \langle \epsilon^{\ast}_2 \rangle$ in region A.
	Neglecting the  $\epsilon_1$-dependence of the threshold $\epsilon_2^{\ast}$ (which is weak on the scale of $\sigma_{E}$ around $\langle \epsilon_1 \rangle$), the convolution with the source parameter distributions given by Eq.~\eqref{eq:genstatistics} gives 
	$\sigma_{t}^{(A)} = (\sigma^{\min}_{t})^2+\sigma^2_{\Delta t}$ where 
	$\sigma^2_{\Delta t} =\sigma_{t1}^2+(1-r_2^2) \sigma_{t2}^2 $, form which the beam splitter contribution $\sigma^{\min}_{t}$, and hence
	$\omega$, can be estimated bypassing the need for a simulation.
	
	\section*{Acknowledgements}
	We acknowledge discussions with Piet W. Brouwer, Jonathan Fletcher, and Masaya Kataoka, and thank Pascal Degiovanni for suggesting the phase shift considerations. E.P., M.K., G.B., and V.K. are supported by grant no.~lzp-2021/1-0232 from the Latvian Council of Science. P.S. and P.R. acknowledge financial support by the Deutsche Forschungsgemeinschaft (DFG, German Research Foundation) within the framework of Germany’s Excellence Strategy-EXC-2123 QuantumFrontiers-390837967. This work was supported in part by the Joint Research Project SEQUOIA (17FUN04) which received funding from the European Metrology Programme for Innovation and Research (EMPIR) cofinanced by the Participating States and from the European Unions Horizon 2020 research and innovation programme.

	\let\oldaddcontentsline\addcontentsline
	\renewcommand{\addcontentsline}[3]{}

	\let\addcontentsline\oldaddcontentsline

	\onecolumngrid
	\newpage
	\section*{Supplementary information}\label{sec:supp}	
	
	\beginsupplement

	\tableofcontents
	\addtocontents{toc}{\protect\setcounter{tocdepth}{2}}
	
	\newpage 
	
	\makeatletter
	\let\@seccntformatorig\@seccntformat
	\def\@seccntformat#1{%
		\ifnum0=\pdfstrcmp{#1}{section}%
		SUPPLEMENTARY FIGURE \csname the#1\endcsname.{} 
		\else
		\@seccntformatorig{#1}%
		\fi
	}
	\makeatother
	
	\section{Micrograph of the sample}\label{sec:supp:micrograph}
	
	\begin{figure*}[htbp]
		\centering
		\includegraphics{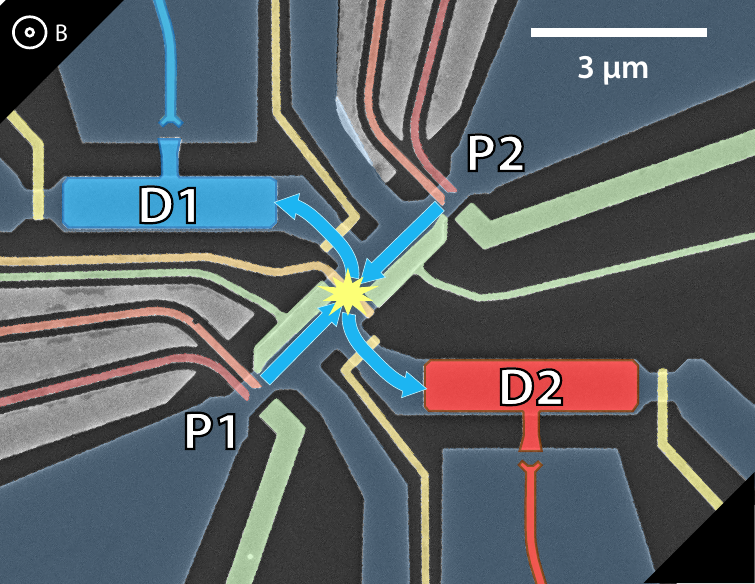}
		\caption{SEM image showing the layout of the CrAu top gates over the etched mesa channel forming the pumps (red), sidewall depletion (green), quantum dot for charge read out and coupling floating gate (blue, red), and entrance/exit to the detector nodes (yellow). The blue arrows indicate the path of the injected electrons.}
		\label{fig:figS1}
	\end{figure*}
	
	\newpage
	
	\section{Additional data for energy exchange}\label{sec:supp:additionaldata}

	\begin{figure*}[htbp]
		\centering
		\includegraphics[width=255pt]{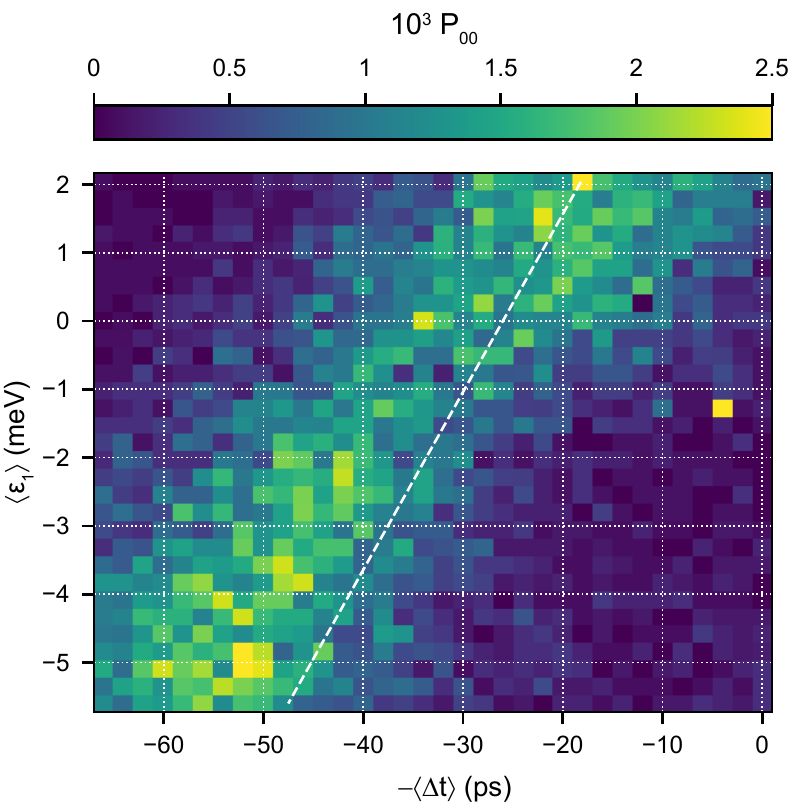}
		\caption{The probability $P_{00}$ that both detectors recorded no electron for the dataset shown in Fig.~\ref{fig:fig3} and Fig.~\ref{fig:fig4}. }
		\label{fig:supP00}
	\end{figure*}


	
	\makeatletter
	\let\@seccntformatorig\@seccntformat
	\def\@seccntformat#1{%
		\ifnum0=\pdfstrcmp{#1}{section}%
		SUPPLEMENTARY NOTE \csname the#1\endcsname.{} 
		\else
		\@seccntformatorig{#1}%
		\fi
	}
	\makeatother
	
	\newpage
	
	\section{Analytic expression for \texorpdfstring{$\Delta \tau$}{Delta tau} for the unscreened Coulomb interaction}\label{sec:app:DeltaTau}
	
	Here we show that
	for the Coulomb potential parametrized as 
	$V_{\text{int}}(r)=U d_0/r$
	the interaction-induced delay time $\Delta \tau$ can be expressed  as
	\begin{multline}
	\exp [ \omega \, \Delta \tau(E)] = |z|   \frac{1+\kappa^2}{2 \,  \kappa^2 \, \xi_m^2(z)} \times \\
	\exp \left \{ 2 \!\!
	\int_{\xi_m(z)}^{\infty} \! \left [  \frac{\kappa \, \xi^2}{(\xi^3 \, \kappa^2  -2 +2 \, \xi \,  z)^{1/2} (\xi^3 \,   +2 -2 \, \xi \,  z)^{1/2} } - \frac{1}{\xi} \right ] d\xi \right \} \label{eq:PhiMartins}
	\end{multline}
	where $z=E/U$, $\kappa=\omega_x/\omega_y$,
	and
	\begin{equation}\label{eq:dmin}
	\xi_m(z) =
	\begin{cases}
	\kappa^{-1} \left[ 
	( \sqrt {(2 z/3 )^{3} +\kappa^{2}} + \kappa  )^{1/3} + ( \kappa - \sqrt {(2   z/3 )^{3} +\kappa^{2}} )^{1/3}
	\right]
	, & z<3/2, \\
	(8z/3)^{1/2}
	\cos \left(
	\arccos\left[\left(2z/3\right)^{-3/2}\right]/3 - \pi / 3 
	\right),
	& z>3/2,
	\end{cases}
	\end{equation}
	with $(\cdot)^{1/3}$  denoting the arithmetical cube root,
	and  the exact minimal distance  reached in Coulomb scattering \cite{Pavlovska2022} is
	\begin{align}
	d_{\text{min}}^{(\alpha=1)}=\xi_m(E/U) \, d_0 \label{eq:dminCoulomb}
	\end{align}
	for 
	$E=\epsilon_1'+\epsilon_2'$.
	
	To formally define  $\Delta \tau $, we first introduce the traversal time $\tau\taudepx{x_0}= \int dx/(\partial \mathcal{H}/\partial p) $, with integral computed along the trajectory $\mathcal{H}(x,y)=E$,  for the relative coordinate to travel from $-x_0 < 0$    to  either $x_0>0$ (in case of transmission of the relative coordinate) or back to $x_0$ (in case of reflection of the relative coordinate), as per \cite[Eq.~(15)]{Pavlovska2022}. Due to the symmetry, 
	$\tau\taudepx{x_0}= 2\int_{-x_0}^0 dx/(\partial \mathcal{H}/\partial p) $ 
	or
	$\tau\taudepx{x_0}= 2\int_{-x_0}^{-d_{\text{min}}} dx/(\partial \mathcal{H}/\partial p) $ (in case of transmission or reflection of the relative coordinate, respectively).
	With $\tau_0\taudepx{x_0}$ standing for $\tau\taudepx{x_0}$ in the absence of interactions, the interaction-induced delay time $\Delta \tau\taudepx{x_0}$ on the described segment of the trajectory is defined as $\tau\taudepx{x_0} - \tau_0\taudepx{x_0}$. Now the total delay time  $\Delta \tau $ is defined as the limit of $\Delta \tau\taudepx{x_0} $ as $x_0 \to +\infty$.
	
	We proceed to prove \eqref{eq:PhiMartins} in three steps: 
	by a change of variables we obtain an expression for $\tau\taudepx{x_0}$; 
	then we characterize the minimal distance reached between the two electrons via \eqref{eq:dmin}; finally, we similarly describe $\tau_0\taudepx{x_0}$ and its asymptotics that allows to find the required limit of $\Delta \tau\taudepx{x_0} $.
	
	\paragraph{Characterization of the traversal time \texorpdfstring{$\tau\taudepx{x_0}$}{tau[x0]}}
	Explicitly, the trajectory  $ \mathcal{H}(x,y)=E$ in the relative coordinates is described by
	\begin{equation}\label{eq:appA:trajectoryCanonical}
	{ y^2-\kappa^2 x^2}{} +\frac{2 d_0^3}{\sqrt{x^2+y^2}} = 2 d_0^2 z,
	\end{equation}
	Introduce new variables $r, \phi$ by $x =d_0 r \cos \phi$, $y = d_0 r \sin \phi$, then \eqref{eq:appA:trajectoryCanonical} becomes
	\begin{equation}\label{eq:appA:trajectoryPolar}
	r^3 \left( \sin^2\phi -  \kappa^2 \cos^2\phi \right)
	- 2  r z +2 = 0.
	\end{equation}
	As $x$ changes along the trajectory from $-x_0$ to either $0$ or $-d_{\text{min}}$, the variable $r$ changes from $r_0$ to $
	d_{\text{min}}/d_0$.
	We express the derivative of $r$ w.r.t. $\tau$   as
	$
	\frac{dr}{\omega d\tau} = 
	\frac{\partial r}{\partial x}  \cdot \frac{dx}{\omega d\tau }
	+
	\frac{\partial r}{\partial y}  \cdot \frac{dy}{\omega d\tau }
	$,
	where the velocity vector $(\frac{dx}{\omega d\tau }, \frac{dy}{\omega d\tau })$ is found as a tangent to the trajectory, arriving at 
	\begin{equation*}
	\frac{dr}{\omega d\tau} =  -\frac{   (\kappa^2+1)r \sin(2\phi) } {2 \kappa } .
	\end{equation*}
	To determine  $\sin (2\phi)$, rewrite \eqref{eq:appA:trajectoryPolar} as $ r^3 \left( \sin^2\phi -  \kappa^2 \cos^2\phi \right)
	= 2  (r z-1)(\sin^2\phi + \cos^2\phi)$
	and express $\tan\phi$,
	\begin{equation*}
	\tan \phi =\pm  \sqrt{\frac{ r^3 \kappa^2  +   r z -2 }{r^3 - 2  r z  +2 }}.
	\end{equation*}
	Here $\pm $ equals the sign of $x \cdot y$ (which is positive before $x$ reaches 0 or $-d_{\text{min}}$, since $x<0$ and $y<0$ until then). The identity $\sin(2 \phi ) =  \frac{2\tan \phi}{\tan^2\phi+1}$ then leads to
	\begin{equation*}
	\frac{\omega d\tau}{dr} 
	= 
	- \frac{ \kappa    r^2}
	{ 
		\sqrt{(r^3  -2  r z+2)(\kappa ^2 r^3 + 2  r z-2)}
	}.
	\end{equation*}
	Consequently, 
	\begin{equation*}
	\omega \tau\taudepx{x_0}/2 = \int_{d_{\text{min}}/d_0}^{r_0} 
	\frac{ \kappa    r^2 \, dr }
	{ 
		\sqrt{(r^3  -2  r z+2)(\kappa ^2 r^3 + 2  r z-2)}
	}.
	\end{equation*}

	\paragraph{Characterization of the minimal distance \texorpdfstring{$d_{\text{min}}$}{d-min}}\label{sec:dMin}
	Furthermore, the value $d_{\text{min}}/d_0$ itself can be found from \eqref{eq:appA:trajectoryPolar}, since the minimal distance is achieved either when $(x,y)= (-d_{\text{min}}, 0)$ or when $(x,y)=(0,-d_{\text{min}})$. In the former case (reflection of the relative coordinate), $z<3/2$ and $\phi=-\pi$; in the latter (transmission of the relative coordinate), $z>3/2$ and $\phi=-\pi/2$. Thus $d_{\text{min}}/d_0=r $ is the solution of $-\kappa^2 r^3 =  2  r z  -2$ or $ r^3 =  2  r z  -2$, respectively.
	
	Consider the case $z>3/2$ and the respective equation $ r^3 - 2  r z  +2 = 0$. The discriminant of this polynomial is $4 \left(8 z^3-27\right)>0$, therefore the equation has three real roots. Vieta's formulas imply that exactly one of them is negative.
	It can be deduced from Vieta's formulas that one of the two positive roots is less than 1 and the other root is larger than one; since $d_{\text{min}}\geq d_0$ in the case of  transmission of the relative coordinate, the relevant root is the largest one. By applying Vieta's trigonometric expression for the cubic equation roots, the three real roots are
	\begin{equation*}
	r^{[k]} = 2 \sqrt{\frac{2z}{3}} \cos \left( 
	\frac{1}{3}\arccos \left( -(2z/3)^{-3/2}  \right) - k \frac{2\pi}{3}
	\right),\quad k \in \{0,1,2\},
	\end{equation*}
	which can be simplified to 
	\begin{equation*}
	r^{[k]} = 2 \sqrt{\frac{2z}{3}} \cos \left( 
	\frac{1}{3}\arccos \left( (2z/3)^{-3/2}  \right) +  \frac{(2k-1)\pi}{3}
	\right),\quad k \in \{0,1,2\},
	\end{equation*}
	with $r^{[0]}$ being the largest of the three roots. 
	Therefore, for $z>3/2$  the value of $ d_{\text{min}}/d_0$ is equal to
	\begin{equation*}
	\sqrt{8z/3} 
	\cos \left(
	\arccos\left[\left(2z/3\right)^{-3/2}\right]/3 - \pi / 3 
	\right).
	\end{equation*}

	Consider the case $z<3/2$ and the respective equation $ \kappa^2 r^3 +  2  r z  -2 =0$. The discriminant of the left-hand side is $-4 \kappa ^2 \left(27 \kappa ^2+8 z^3\right)$, which is negative for $ z> z_*:=-\frac{3}{2}  \kappa ^{2/3} $  and positive for $z< z_*$. Hence, $r$ is the unique real (and positive) solution for $z \in (z_*, 3/2)$. When $z< z_*$, there are three distinct real roots; Vieta's formulas imply that exactly one of them is positive, thus $r$ is the unique positive solution. In both cases, it follows from Cardano's formula  that 
	the positive solution is
	\begin{equation*}
	\frac{( \, \sqrt {(2 \, z/3 )^{3} +\kappa^{2}} + \kappa \, )^{1/3} + ( \kappa - \sqrt {(2 \, z/3 )^{3} +\kappa^{2}} \,)^{1/3}}{\kappa}.
	\end{equation*}
	This completes the proof that $ d_{\text{min}}/d_0$  is given by the function $\xi_m(z)$, defined in \eqref{eq:dmin}.

	\paragraph{Characterization of the traversal time \texorpdfstring{$\tau_0\taudepx{x_0}$}{tau0[x0]}}
	In the absence of interactions,    
	the trajectory in the relative coordinates is described by
	\begin{equation}\label{eq:appA:trajectory0Canonical}
	{ y^2-\kappa^2 x^2}{} = 2 d_0^2 z.
	\end{equation}
	Similarly as we did for $\omega \tau\taudepx{x_0}$,  one can   derive
	\begin{align*}
	\omega \tau_0\taudepx{x_0}
	& = 
	2\int_{\rho_m}^{\rho_0} \frac{\kappa  r \, dr }{\sqrt{r^2-2 z} \sqrt{\kappa ^2 r^2+2 z}} 
	= 
	\left.
	-2\ln\left( \sqrt{\kappa ^2 r^2+2 z}-\kappa  \sqrt{r^2-2 z} \right) \right\vert_{r=\rho_m}^{r=\rho_0},
	\end{align*}
	where $\rho_0 = \sqrt{x_0^2+y_0^2}/d_0$, with $y_0$ determined by \eqref{eq:appA:trajectory0Canonical}, 
	and the minimal distance reached between the two electrons now is
	\begin{equation*}
	\rho_m = 
	\begin{cases}
	\sqrt{2z}, & z >0, \\
	\sqrt{-2z} / \kappa,  & z<0.
	\end{cases}
	\end{equation*}

	Both for $z>0$ and $z<0$ one obtains
	\begin{equation*}
	\omega \tau_0\taudepx{x_0}
	=
	\ln  \left(2 \left(\kappa ^2+1\right) \vert z\vert \right)
	-2 \ln \left\vert \sqrt{\kappa^2 \rho_0^2 + 2z }-   \kappa\sqrt{\rho_0^2-2 z}\right\vert .
	\end{equation*}
	Furthermore, since 
	\begin{equation*}
	\lim_{\rho\to \infty } \rho \left(
	\sqrt{2 z+\rho^2 \kappa ^2}
	-\kappa\sqrt{\rho^2-2 z}
	\right)= \frac{z(\kappa^2+1)}{\kappa},
	\end{equation*}
	we have
	\begin{equation*}
	-2 \ln \left\vert \sqrt{\kappa^2 \rho_0^2 + 2z }-   \kappa\sqrt{\rho_0^2-2 z}\right\vert 
	=
	\ln \frac{\kappa^2}{z^2(\kappa^2+1)^2} + 2 \ln \rho_0 + o(\rho_0)
	\end{equation*}
	as $\rho_0 \to \infty$. This gives the following asymptotics for large $\rho_0$ (equivalently, large $x_0$):
	\begin{equation*}
	\omega \tau_0\taudepx{x_0}
	=
	\ln \frac{2\kappa^2 \rho_0^2}{\vert z\vert  (\kappa^2+1)}   + o(x_0).
	\end{equation*}
	By expressing $\ln r_0 = \ln \xi_m + \int_{\xi_m}^{r_0} \frac{d\xi}{\xi}  $ we now see that
	\begin{align*}
	\omega \Delta \tau\taudepx{x_0} 
	& =  \omega  \tau\taudepx{x_0}  -  \omega  \tau_0\taudepx{x_0} 
	\\
	&  = 2\int_{\xi_m}^{r_0} 
	\frac{ \kappa    \xi^2 \, d\xi }
	{ 
		\sqrt{(\xi^3  -2  \xi z+2)(\kappa ^2 \xi^3 + 2  \xi z-2)}
	} 
	-
	\ln \frac{2\kappa^2 \rho_0^2} {\vert z\vert  (\kappa^2+1)} + o(x_0)
	\\
	&  = 2\int_{\xi_m}^{r_0} 
	\frac{ \kappa    \xi^2 \, d\xi }
	{ 
		\sqrt{(\xi^3  -2  \xi z+2)(\kappa ^2 \xi^3 + 2  \xi z-2)}
	} 
	-2\int_{\xi_m}^{r_0} \frac{d\xi}{\xi} 
	-
	\ln \frac{2\kappa^2 \rho_0^2 \xi_m^2} {\vert z\vert  (\kappa^2+1) r_0^2}
	+ o(x_0)
	\\
	& =
	\ln \frac{\vert z\vert  (\kappa^2+1)} {2\kappa^2 \xi_m^2}
	+
	2\int_{\xi_m}^{r_0} \left[
	\frac{ \kappa    \xi^2   }
	{ 
		\sqrt{(\xi^3  -2  \xi z+2)(\kappa ^2 \xi^3 + 2  \xi z-2)}
	} - \frac{1}{\xi}
	\right] d\xi  + o(x_0).
	\end{align*}
	The last equality is justified by the fact that from \eqref{eq:appA:trajectoryCanonical} and \eqref{eq:appA:trajectory0Canonical} it follows that
	\begin{align*}
	\frac{r_0^2}{\rho_0^2}
	& = \frac{x_0^2(\kappa^2+1) -\frac{2 d_0^3 }{r_0^2}  + 2d_0^2 z}{x_0^2(\kappa^2+1)   + 2d_0^2 z} 
	= 1 - \frac{2 d_0^3 }{r_0^2 \rho_0^2 }  
	=  1-o(x_0).
	\end{align*}
	Taking the limit $x_0 \to \infty$, we arrive at \eqref{eq:PhiMartins}.
	Finally, we note that by the limit comparison test  the improper integral 
	$\lim\limits_{r_0\to +\infty}\int_{\xi_m}^{r_0} \left[
	\frac{ \kappa    \xi^2   }
	{ 
		\sqrt{(\xi^3  -2  \xi z+2)(\kappa ^2 \xi^3 + 2  \xi z-2)}
	} - \frac{1}{\xi}
	\right] d\xi $ converges, since 
	\begin{equation*}
	\frac{ \kappa    \xi^2   }
	{ 
		\sqrt{(\xi^3  -2  \xi z+2)(\kappa ^2 \xi^3 + 2  \xi z-2)}
	} - \frac{1}{\xi}
	\sim 
	\begin{cases}
	z(\kappa^2-1)/\kappa^2  \xi^{-3}, & \kappa \neq 1,\, z\neq 0,\\
	-(\kappa^2-1)/\kappa^2 \xi^{-4}, & \kappa \neq 1, \,  z = 0, \\ 
	2z^2 \xi^{-5}, & \kappa =1, \,  z\neq 0, \\ 
	2\xi^{-7}, & \kappa =1, \,  z=0.
	\end{cases}
	\end{equation*}

	\section{Relation between the shifted transmission threshold \texorpdfstring{$\epsilon^{\ast}$}{epsilon*} and the modified relative traversal time \texorpdfstring{$\Delta \tau$}{Delta tau}\label{supp:thresholdtau}}
	
	Condition for either electron being on the threshold is derived in Ref.~\cite{Pavlovska2022} to be
	\begin{align}
	E_{-} \,  \Phi_{\kappa} = \pm E_{+} \,
	\label{eq:ph_boundaries}
	\end{align}
	where in our notation $E_{\pm}=\epsilon_2'\pm\epsilon_1'$ and $\Phi_{\kappa}=\sgn(E_{+}) \exp [ \omega \, \Delta \tau(E_{+}) ]$. The equation is valid with both signs and encodes both $\epsilon_2^{\ast}(\epsilon_1)$ and $\epsilon_1^{\ast}(\epsilon_2)$. 
	For the function $ \epsilon^{\ast}(\epsilon)$ defining the symmetric ($\Delta t=0$) shifted transmission threshold,  we  consider  $\epsilon_2'=\epsilon_2=\epsilon^{\ast}$ and $\epsilon_1'=\epsilon_1=\epsilon$, hence $E_{\pm}= \epsilon^{\ast} \pm \epsilon$. 
	$\Phi_{\kappa}$ as function of $E_{+}$ changes sign at $E_{+}=0$ ($\Delta \tau \to -\infty$) and diverges at $E_{+}=3U/2$ ($\Delta \tau \to +\infty$). For the solution branch defining $\epsilon^{\ast}>0$ as function of $\epsilon$ one has to pick  the following signs in the corresponding three ranges of $E_{+}$:
	\begin{align}
	\begin{cases} 
	(\epsilon^{\ast} - \epsilon) \,  \exp [ \omega \, \Delta \tau ] = -  (\epsilon^{\ast} + \epsilon) \, , & E_{+}<0  \, ,\\ 
	(\epsilon^{\ast} - \epsilon) \,  \exp [ \omega \, \Delta \tau ] =  (\epsilon^{\ast} + \epsilon) \,, & 0< E_{+}<1.5 \, U \, ,\\
	(\epsilon^{\ast} - \epsilon) \,  \exp [ \omega \, \Delta \tau ] = - (\epsilon^{\ast} + \epsilon) \,, & E_{+}>1.5  \, U \, .
	\end{cases} \label{eq:thr}
	\end{align}
	The solutions are $\epsilon^{\ast}/\epsilon = \tanh [\omega \, \Delta \tau/2]$ ($E_{+}<0$ and $E_{+}>1.5\,U$) and $\epsilon^{\ast}/\epsilon = \coth [\omega \, \Delta \tau/2]$ ($0<E_{+}<1.5\,U$).
	As $\epsilon^{\ast}=-\epsilon$ at $E_{+}=0$ and $\epsilon^{\ast}=\epsilon=3 U/4$ at $E_{+}=3 U/2$, the results can be expressed as equation \eqref{eq:threshold} in the main text.

	\section{Cut-off implementation}\label{sec:supp:cutoffimplem}
	\begin{figure*}[htbp]
		\centering
		\includegraphics[width=0.5\textwidth]{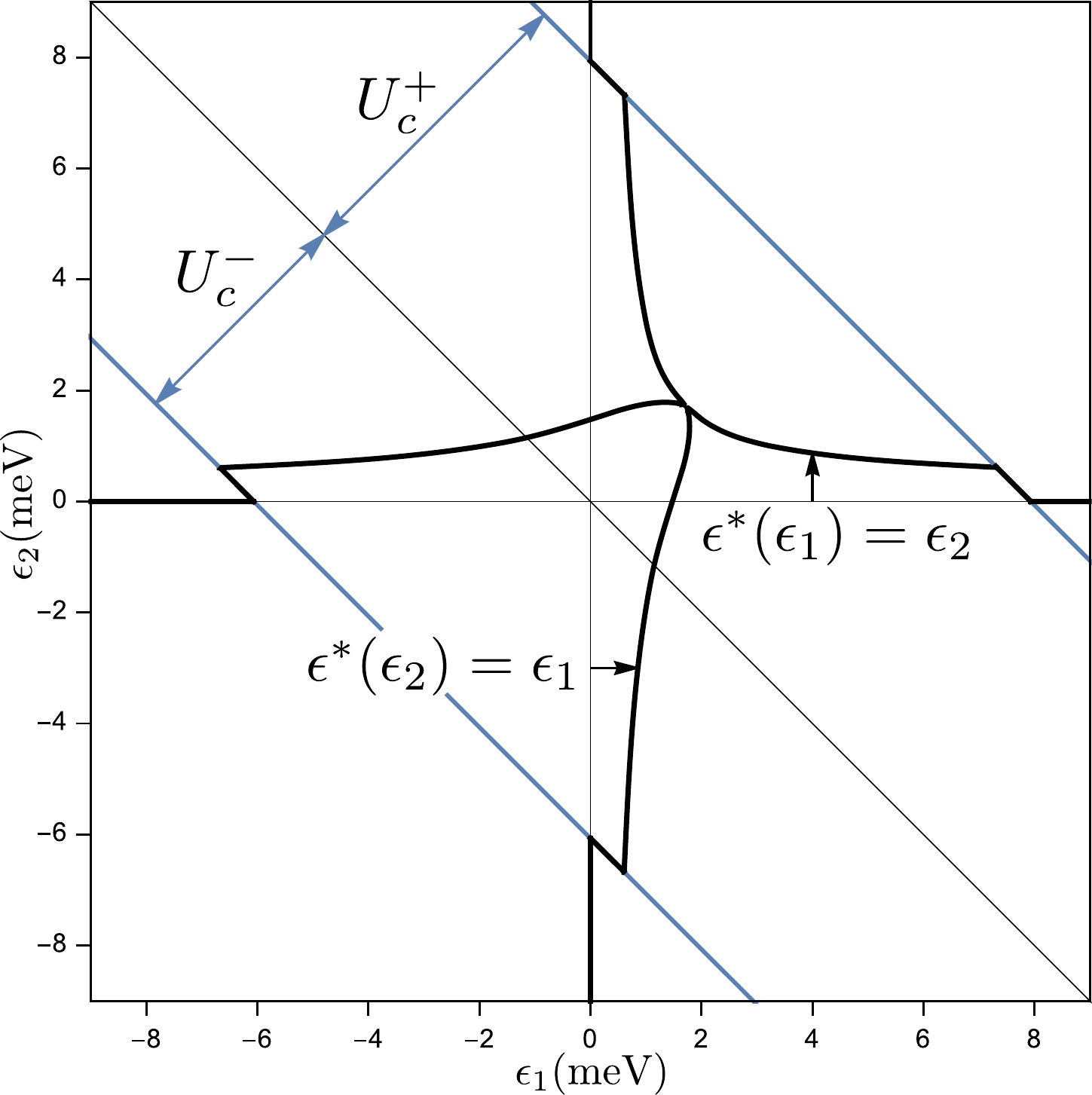}
		\caption{Illustration of the cut-off procedure used in the simulation in terms of the shifted transmission thresholds at $\Delta t=0$. Here $\kappa = 1$, $U=\SI{2.3}{meV}$, $U_{c}=\SI{7.0}{meV}$, corresponding to the simulation parameters used for Fig.~\ref{fig:fig3} in the main text.}
		\label{fig:figST1}
	\end{figure*}
	Two physical effects limit applicability of the exactly solvable analytical model for which Eqs.~\eqref{eq:PhiMartins} and Eq.~\eqref{eq:dmin} apply: finite extent of the beam splitter where the quadratic saddle potential is applicable and the screening of two-electron interaction at large distances. We account for these limitations by a single condition: if the two electrons are predicted within the full model to reach a minimal distance $d_{\text{min}}$ that is less than a fixed cut-off distance $d_{\text{cutoff}}$, the full model is applied. If, however, $d_{\text{min}}>d_{\text{cutoff}}$ then the electrons are treated as non-interacting and the baseline transmission thresholds, $\epsilon^{\ast}_1=\epsilon^{\ast}_2=0$ are applied. This approach underestimates the effect of interactions outside of the beam splitter region, but has the advantage of being compatible with the $\Delta t$ scaling relations which also prevents unphysical extrapolation of exponential acceleration due the quadratic roll-off of the saddle potential for large $|\Delta t|$. 
	
	We apply the above cut-off procedure by defining $U_{c}=m \omega _{x}^{2} d_{\text{cutoff}}^{2} /4$ and expressing the cutoff values of the conserved integral of motion for the relative coordinate, 
	\begin{align}
	U_{c}^{-} & =\mathcal{H}(-d_{\text{cutoff}},0)=-U_{c} + \frac{U^{3/2}\kappa}{\sqrt{2 U_{c}}} \label{eq:Ucminus}\\
	U_{c}^{+} & =\mathcal{H}(0,-d_{\text{cutoff}})=\frac{U_{c}}{\kappa ^2} + \frac{U^{3/2}\kappa}{\sqrt{2 U_{c}}}
	\label{eq:Ucplus}
	\end{align}
	Note that $d_{\text{cutoff}}$ as a function of $U_{c}^{\pm}$ in the Eq.~\eqref{eq:Ucminus} and \eqref{eq:Ucplus} equals $d_{\min}$ as a function of $E$ in the equation \eqref{eq:dminCoulomb} and the cut-off condition $d_{\text{min}}>d_{\text{cutoff}}$ on $E$ is implemented as as:
	$E > U_c^{+}(d_{\text{cutoff}})$ or 
	$E < U_c^{-}(d_{\text{cutoff}})$.
	
	Figure~\ref{fig:figST1} illustrates the shifted transmission thresholds that correspond to this procedure.

	\section{Power-law interaction potential and first-principles estimates}\label{sec:supp:powerlawpotential}
	Parametrizing a power-law interaction potential as
	\begin{align}
	V_{\text{int}}(r) = \frac{3\,U}{2+\alpha} \, \left (\frac{d_0}{r}\right)^{\alpha} \label{eq:powerlawVint}
	\end{align}
	ensures that the minimum of $\mathcal{H}(0,y)$ as function of $y$ is reached at $y=\pm d_0$ and $\mathcal{H}=3 U/2$ where $d_0$ satisfies $m \omega_y^2 d_0^2/2=\alpha \, V_{\text{int}}(d_0)$, which in turn implies that $\Delta \tau(E)$ diverges as $E \to 3 U/2$ (setting the crossing point of the shifted thresholds to $\epsilon_{1c}+\epsilon_{2c}=3 U/2$) and makes $d_0$ the closest possible distance between the electrons passing each other at the beam splitter, in consistency with 
	the definitions of $U$ and $d_0$ introduced for the special case of $\alpha=1$ in Ref.~\cite{Pavlovska2022}. 
	
	Explicitly,
	\begin{align} \label{eq:d0val}
	d_0 =\sqrt{\frac{6 \alpha}{2 +\alpha}} \sqrt{\frac{U}{m \omega_y^2}}
	\end{align}
	
	Unscreened Coulomb potential in GaAs~\cite{Reimann2002} can be quantified in terms of 
	effective hartree $\hbar \omega^{\star} =\SI{12}{meV}$ and effective Bohr radius $a^{\star}=\SI{9.8}{nm}$ as 
	$V_{\text{int}}(r) = \hbar \omega^{\star} \, a^{\star}/r$, which gives, taking into account that $\hbar/m = \omega^{\star} (a^{\star})^2$, 
	\begin{align}
	U & = 2^{-1/3 }\hbar \left ( \omega^{\star} \, \omega_y^2 \right)^{1/3}  \label{eq:Uvalue} \\
	d_0 &= 2^{1/3} \, a^{\star} \, (\omega^{\star}/ \omega_y)^{2/3}  \label{eq:d0}
	\end{align}
	
	From $\omega_x/\omega_y=1$, 
	$\omega =\omega_x \, \omega_y/\omega_c = \SI{250}{GHz}$ inferred from simulations and $\hbar \omega_c=\hbar e B/m= \SI{17.5}{meV}$, we get $\hbar \omega_y=\SI{1.7}{meV}$ and, using equation \eqref{eq:Uvalue} above, $U=\SI{2.6}{meV}$ which is compared to $U=\SI{2.3}{meV}$ obtained by adjusting $U$ in the simulations. From Eq.~\eqref{eq:d0} we get $d_0=\SI{45}{nm}$.
	
	From $U_c=\SI{7}{meV}$ we get $d_{\text{cutoff}}=\SI{75}{nm}$.
	

	\section{Perturbation theory in the interaction strength}
	\label{sup:interaction_strength}
	
	Applying classical  perturbation theory  to the definition of $\tau(E)$ to the first order in $V_{\text{int}}(r)$ gives 
	\begin{align} \label{eq:perturbativeprimary}
	\omega \Delta \tau(E) =E^{-1} \kappa_{\pm}
	\int_{1}^{\infty} \left [ \frac{(1+\kappa_{\pm}^2) \,\xi\, V_{\text{int}}(\xi d_{\text{min}}) }{\sqrt{\xi^2-1}(\kappa_{\pm}^2 \xi^2+1)^{3/2}} - \frac{ d_{\text{min}} \, V_{\text{int}}'(\xi d_{\text{min}}) }{\sqrt{\xi^2-1}\sqrt{\kappa_{\pm}^2 \xi^2+1}}
	\right ] d \xi
	\end{align}
	where $\kappa_{+}=\omega_x/\omega_y=1/\kappa_{-}$,
	$\pm = \sgn E$,  $V_{\text{int}}'(r)=dV_{\text{int}}(r)/dr$,
	and  
	\begin{align} \label{eq:dmin0}
	d_{\text{min}}^{(U=0)} =  \sqrt{4 |E| \kappa_{\pm}/(m \omega_{x} \omega_{y})}
	\end{align}
	(which is consistent with $U \to 0$ limit of Eq.~\eqref{eq:dminCoulomb}).  
	
	The two terms in the integrand in Eq.~\eqref{eq:perturbativeprimary} account for the interaction-induced  change of velocity and the change of trajectory, respectively.

	The quantum phase shift $\Delta \varphi(E)$ along the non-perturbed trajectory is expressed similarly
	\begin{align}
	\Delta \varphi(E)= (\hbar \omega)^{-1} \int_{-\infty}^{+\infty} V_{\text{int}}[r(t)] dt = \frac{2 \, \kappa_{\pm}}{\hbar \omega}  \int_1^{\infty} \frac{ \xi \, V_{\text{int}}(\xi d_{\text{min}}) \, d\xi }{\sqrt{\xi^2-1}\sqrt{\kappa_{\pm}^2 \xi^2+1}}
	\label{eq:phiparturbative}
	\end{align}
	

	For the scale-free power law potential \eqref{eq:powerlawVint}, the integrals in the linear expansion
	\eqref{eq:perturbativeprimary} can be computed exactly (regardless of the prefactor notation)
	\begin{align} \label{eq:DeltaTau}
	\omega  \, \Delta \tau(E) \, E=c(\kappa_{\pm},\alpha) \, V_{\text{int}}(d_{\text{min}}) 
	\end{align}
	where the dimensionless coefficient is
	\begin{align}
	c(\kappa,\alpha)& =\kappa  \int_{1}^{\infty} \frac{ \xi^{-1-\alpha} \left [ \alpha +
		\left ( 1+ \kappa^2 +\alpha \, \kappa^2 \right) \xi^2 \right ]}{{\sqrt{\xi^2-1}(\kappa^2 \xi^2+1)^{3/2}}} d \xi  =  \int_{1}^{\infty} \frac{ \alpha \, \kappa \, \xi^{1-\alpha} \, d \xi }{\sqrt{\xi^2-1}\sqrt{\kappa^2 \xi^2+1}} \label{eq:magic} \\
	& =  \frac{\sqrt{\pi} \, \Gamma\left (1+\frac{\alpha}{2}\right)}{\Gamma\left(\frac{1+\alpha}{2}\right)} {}_2F_1\left( \frac{1}{2} , \frac{\alpha}{2}; \frac{1+\alpha}{2}; - \frac{1}{\kappa^2} 
	\right)  \\
	c(1,\alpha)& =\sqrt{\pi} \, \frac{\Gamma\left(1+\frac{\alpha}{4}\right)}{\Gamma\left(\frac{1}{2}+\frac{\alpha}{4}\right)} =\begin{cases}
	1 + \frac{(\ln 2) \alpha}{2} , & \alpha \ll 1 \\
	\frac{\sqrt{\alpha \pi}}{2} , &  \alpha \gg 1
	\end{cases} \\
	c(\kappa, 1) & = K(-1/\kappa^2) = \begin{cases}
	\kappa \ln \frac{4}{\kappa} , & \kappa \ll 1 \\
	\frac{\pi}{2} -\frac{\pi}{8 \kappa^2} , &  \kappa \gg 1 
	\end{cases} 
	\end{align}
	where $\Gamma$ is the Gamma function, ${}_2F_1$ is the hypergeometric function,  and $K$ is the complete elliptic integral of the first kind.
	
	As explained in the main text, in the small $\Delta \tau$ limit, Eq.~\eqref{eq:DeltaTau} with $E=\epsilon$ gives directly the threshold function 
	\begin{align}
	\epsilon^{\ast}(\epsilon) = 2 c(\kappa_{\pm},\alpha) V_{\text{int}}[ d_{\text{min}} (\epsilon)] \ll |\epsilon| \label{eq:thresholdpert}
	\end{align}
	Using the prefactor defined in Eq.~\eqref{eq:powerlawVint} with Eq.~\eqref{eq:dmin0} gives the large-$|\epsilon|$ asymptotics of the shifted thresholds:
	\begin{align}
	\epsilon^{\ast}(\epsilon) = U \, (U/|\epsilon|)^{-\alpha/2} \, 6 \,  \left( \frac{3 \alpha}{2} \right )^{\alpha/2} (2+\alpha)^{-1-\alpha/2} \times 
	\begin{cases}
	c(\kappa, \alpha) \,,  & \epsilon>0 \\
	k^{\alpha} c(\kappa^{-1}, \alpha) & \epsilon<0   
	\end{cases}
	\end{align}

	For the quantum phase estimate \eqref{eq:phiparturbative} and the power-law interaction \eqref{eq:powerlawVint},  we find with the help of Eq.~\eqref{eq:magic}
	\begin{align}
	\Delta \varphi(E) = \frac{2 \, c(\kappa_{\pm},\alpha)}{\alpha} \frac{  V_{\text{int}}(d_{\text{min}})  }{\hbar \omega} 
	\end{align}
	With $E=\epsilon/2+\epsilon/2$ and  Eq.~\eqref{eq:thresholdpert} this proves the relation  $\Delta \varphi(\epsilon)=\alpha^{-1} \epsilon^{\ast}(\epsilon) /(\hbar \omega) $   quoted in the main text for the experimentally relevant case of the Coulomb potential $\alpha=1$.
	
	Consistency  of using Eq.~\eqref{eq:phiparturbative}, and, subsequently Eq~\eqref{eq:figureotmerit} of the main text, requires localization of the quantum wave-packet in time domain
	within the beam splitter time-scale,  $\sigma_t < \omega^{-1}$,
	and the no-reflection condition $|\epsilon/2|\gg\sigma_E$.
	Due to energy-time uncertainty $\sigma_E \sigma_t >\hbar/2$ this requires
	$|\epsilon| \gg \hbar \omega$  which is compatible with
	$\Delta \varphi$ of order one as
	$\hbar \omega \sim \epsilon^{\ast}(\epsilon) \ll |\epsilon|$ is possible in the perturbative regime for large $U \gg \hbar \omega$.
	
	\section{Energy exchange and electron loss calculation}
	\label{sup:energy_exchange}
	
	The motion of the two electrons in the quadratic saddle potential can be fully separated into relative and center-of-mass coordinates that implies not only the conservation of the true asymptotic energy $\epsilon_1+\epsilon_2$ but also of the integral of motion $E=\epsilon_1'+\epsilon_2'$ associated with  the relative coordinate~\cite{Pavlovska2022}. These two conservation laws allow one to express the relation between the incoming initial ($\epsilon_1$, $\epsilon_2$, $\Delta t$) 
	and the outgoing final ($\epsilon_1^{f}$, $\epsilon_2^{f}$, $\Delta t^{f}$) non-interacting parameters via a single function,
	$\Phi_{\kappa}=\sgn(E) \exp [ \omega \, \Delta \tau(E) ]$, for which we provide a  closed form expression, Eq.~\eqref{eq:PhiMartins}. The relation for $\Delta t^{f}$ is derived in Ref.~\cite{Pavlovska2022}. Similarly, we derive the expression for the energy transfer, 
	\begin{equation}
	\epsilon_2^{f}-\epsilon_2=-(\epsilon_1^{f}-\epsilon_1) =
	\frac{\pm 1 + \Phi_{\kappa}  }{2 \, \Phi_{\kappa}} 
	\left [ \mp (\epsilon_2' - \epsilon_1') \, \Phi_{\kappa}  +E \tanh \frac{\omega \Delta t}{2} \right] \, ,
	\end{equation}
	where the upper sign is for $E< 3U/2$.
	Note that the energy exchange is singular at $E=3U/2$ as at this combination of initial conditions the electrons stay together ($\Delta t^{f}=0$) indefinitely within the classical solution. The real amount of the energy exchange is limited by the finite extent of the beam splitter but may exceed both $|\epsilon_1|$ and  $|\epsilon_2|$ significantly, leading to the distinct loss signal.

	The losses in each channel are calculated using a simple model for loss rate energy dependence $P^{\text{loss}}_{1,2}=1-\exp \left [-C_{1,2} \exp (-\beta_{0} \,  \epsilon_{1,2}^{f}) \right ]$ at  detector D1 or D2, respectively. The parameters  $C_{1}=0.014$, $C_{2}=0.004$ and $\beta_{0}=\SI{0.25}{meV^{-1}}$ are adjusted to reproduce the absolute contrast of the the loss signals $P_{10}$ and $P_{01}$ while all the other parameters of collision kinematics and statistical broadening are fixed by the  simulation of the correlation signatures in the lossless collision outcomes, $P_{20}$ and $P_{02}$.

	\ifdefined\drafting
	\newpage
	\else
	\fi
	
	
\end{document}